\documentclass[lettersize,journal]{IEEEtran}
\usepackage{amsmath,amsfonts}
\usepackage{algorithm}
\usepackage{algorithmicx}
\usepackage{enumitem}
\usepackage{array}
\usepackage[caption=false,font=normalsize,labelfont=bf]{subfig}
\usepackage{textcomp}
\usepackage{stfloats}
\usepackage{url}
\usepackage{verbatim}
\usepackage{graphicx}
\usepackage{cite}
\usepackage{color}
\usepackage{algpseudocode}
\usepackage[font=normalsize,labelfont=bf]{caption}
\title{Generalization of Deep Reinforcement Learning for Jammer-Resilient Frequency and Power Allocation
}

\author{Swatantra Kafle, Jithin Jagannath, Zackary Kane, Noor Biswas, Prem Sagar Vasanth Kumar, Anu Jagannath
\vspace{-0.8 cm}
\thanks{Swatantra Kafle, Jithin Jagannath, Zackary Kane, Noor Biswas, Prem Sagar Vasanth Kumar, and Anu Jagannath are with the Marconi-Rosenblatt AI/ML Innovation Laboratory, ANDRO Computational Solutions, LLC, 7980 Turin Rd, Rome, NY 13440, USA. Emails: \{skafle, jjagannath, zkane, nbiswas, pkumar, ajagannath\}@androcs.com
}}
\begin{document}

\maketitle
%\thanks{Anu Jagannath is with Northeastern University and Marconi-Rosenblatt AI/ML Innovation Laboratory, ANDRO Computational Solutions, LLC.}

\begin{abstract}
We tackle the problem of joint frequency and power allocation while emphasizing the generalization capability of a deep reinforcement learning model. Most of the existing methods solve reinforcement learning-based wireless problems for a specific pre-determined wireless network scenario. 
The performance of a trained agent tends to be very specific to the network and deteriorates when used in a different network operating scenario (e.g., different in size, neighborhood, and mobility, among others). We demonstrate our approach to enhance training to enable a higher generalization capability during inference of the deployed model in a distributed multi-agent setting in a hostile jamming environment. 
With all these, we show the improved training and inference performance of the proposed methods when tested on previously unseen simulated wireless networks of different sizes and architectures. More importantly, to prove practical impact, the end-to-end solution was implemented on the embedded software-defined radio and validated using over-the-air evaluation.
\end{abstract}


\section{Introduction and Background}

The dramatic increase in the number of connected wireless devices with the demand for higher data rates has demanded increasingly efficient use of wireless resources. Modeling next-generation of complex wireless systems and the dynamic allocation of scarce resources demand data-driven methodologies such as powerful function approximation used by deep reinforcement learning (DRL). In the last decade, several classes of machine learning algorithms were developed that have expedited research and development in different domains, including the wireless networks \cite{jagannath2021redefining,   li2018intelligent,  yu2019deep, jagannath2020neural,  naparstek2018deep,jagannath2019machine}. DRL has been studied in the wireless community, especially for the problem of dynamic resource management \cite{maaz2017energy, anandkumar2011distributed, nasir2020deep, modi2019transfer, ahmed2019deep, nasir2019multi}. 
Power control and frequency (channel) selection have been studied in both independent and joint settings. Authors in \cite{ahmed2019deep} developed a centralized deep Q-Network (DQN)-based algorithm for downlink power control. The work in \cite{nasir2019multi} developed a deep distributed approach for multi-agent reinforcement learning (MARL) for power control to maximize the network weighted sum rate where each agent (transmitter) exchanges its instantaneous observation with its nearby transmitter. 
The work in \cite{tan2020deep} assumes Device-to-Device (D2D) networks and addresses the problem of joint power control and frequency allocation. In the problem setup, a macro base station (MBS) provides signaling for the synchronization of D2D pairs and assists in allocating pilots. In \cite{nasir2021deep}, a DRL-based joint power control and channel selection solution is developed in a cellular network setup.
% {In \cite{li2018intelligent}, a DRL-based power control algorithm is developed in a cognitive network setup, where a secondary user exploits DRL to learn to transmit policy based on the power control policy of the primary user such that quality of service (QoS) is guaranteed at both primary users and secondary users.}
% The work in \cite{ghadimi2017reinforcement} deals with the problem of power control and rate adaptation to optimize a network-wide utility function in the MARL setting.
In contrast, \cite{jagannath2022mr} considers distributed multiple transmitter-receiver pairs (Tx-Rx pairs) setup making power control decisions but lacks the joint frequency selection considered in this work.  

% In most cases, these works are limited to simulation and do not consider real-world conditions. For example, the agents are trained on a specific network, and when these agents are deployed in a different network structure, the performance deteriorates.   

% {\color{blue} In most cases, these works are limited to simulation and do not generalize. We say a trained agent generalizes if the agent has robust inference performance when deployed in different network operating scenarios such as size, neighborhood, mobility, and jamming profile. The idea of generalization does not extend to the system parameters of the radio, such as the number of available channels and the power level, i.e., these parameters are assumed to be prespecified and continue to be consistent during the deployment.}

In most cases, these works are limited to simulation and do not generalize. We define the generalization capability of an agent as the ability of the trained agent to provide robust performance at inference when deployed in different (unseen during training) operating environment parameters such as network size, neighborhood, mobility, and jamming profile. While these are not a comprehensive list of operating environment parameters, we expect them to be defined by the operator or use case. The idea of generalization does not extend to the system parameters of the radio (that usually represent actions), such as the number of available channels and the power level, i.e., these parameters are assumed to be prespecified and continue to be the same during the deployment.
None of these works address the generalization capability of agent performance during inference in networks that have mismatches to the training environment. 
In real-world scenarios, the wireless network used for training and testing will most likely differ,  resulting in performance degradation. 
So, trying to incorporate all the specifics of radio frequency (RF) environments and operating scenarios will lead to explosions of the solution space. This results in an exponential increase in sample complexity, demanding very high computational resources. 
We address these problems by proposing a novel training approach to the multi-agent problem in real-world wireless networks. To demonstrate the effectiveness of our approach, we consider a problem of joint power control and channel selection among the wireless nodes in an active jamming environment in a low probability of intercept and detection (LPI/D) networks due to its relevance to a tactical wireless network. To the best of our knowledge, there is no previous work that has designed and deployed DRL for joint power and channel selection in a hostile jamming environment on actual Software defined radio (SDR). We use direct-sequence code-division multiple access (DS-CDMA) to implement a robust LPI/D physical layer due to its advantages, such as easy frequency management and low peak-to-average power ratio (PAPR), among others. We empirically show the convergence of the proposed training approach. However, analysis of theoretical convergence in multi-agent RL problems with model parameter averaging when the global state space is partially observable in each agent is still an open problem in the literature \cite{zhang2021multi, qi2021federated }.

\textbf{Contribution:} In this work, (i) we develop a new training approach to enhance the generalization of multi-agent DRL problems in wireless networks. We demonstrate the effectiveness of the solution by addressing the problem of joint power control and frequency selection in an active jamming environment on actual radio hardware for the first time in literature (to the best of our knowledge). (iii) We provide extensive results of the proposed training method when the trained agent is deployed in simulated and actual wireless networks of different sizes and architectures. (iii) Furthermore, the solution is validated using over-the-air (OTA) evaluation when the DRL agents are deployed at the edge on embedded SDRs. (iv) Finally, we provide a video demonstration of the solution operating in real-time in real-world outdoor deployment.

\section{Problem Formulation and Proposed Solution}

Consider a wireless network consisting of several transceiver pairs using DS CDMA protocol. Assume that there are $2K$ wireless transceivers, each with spreading sequence matrix $\mathbf S \in \mathbb C^{L \times 2K}$, where $L$ is the length of the spreading code. This forms $K$ transmit-receive pairs that can transmit at different power levels, $p_1 \cdots, p_{n_p}$, and multiple frequencies $f_1,\cdots, f_{n_f}$, where $n_p$ and $n_f$ are the numbers of power level and available frequencies, respectively. 
All agents learn policy to choose power level and channel by maximizing the same discounted sum of rewards. We assume each agent has partial observability, i.e., it cannot observe the entire underlying Markov state. Hence, we are interested in designing a reward function such that agents prefer to transmit at the lowest power and use available channels uniformly when maximized. We focus on the maximization of rewards through some form of centralized training. In this setup, each agent can share their learned model for aggregation. 

\begin{figure} 
  \centering
    	\includegraphics[width=8cm]{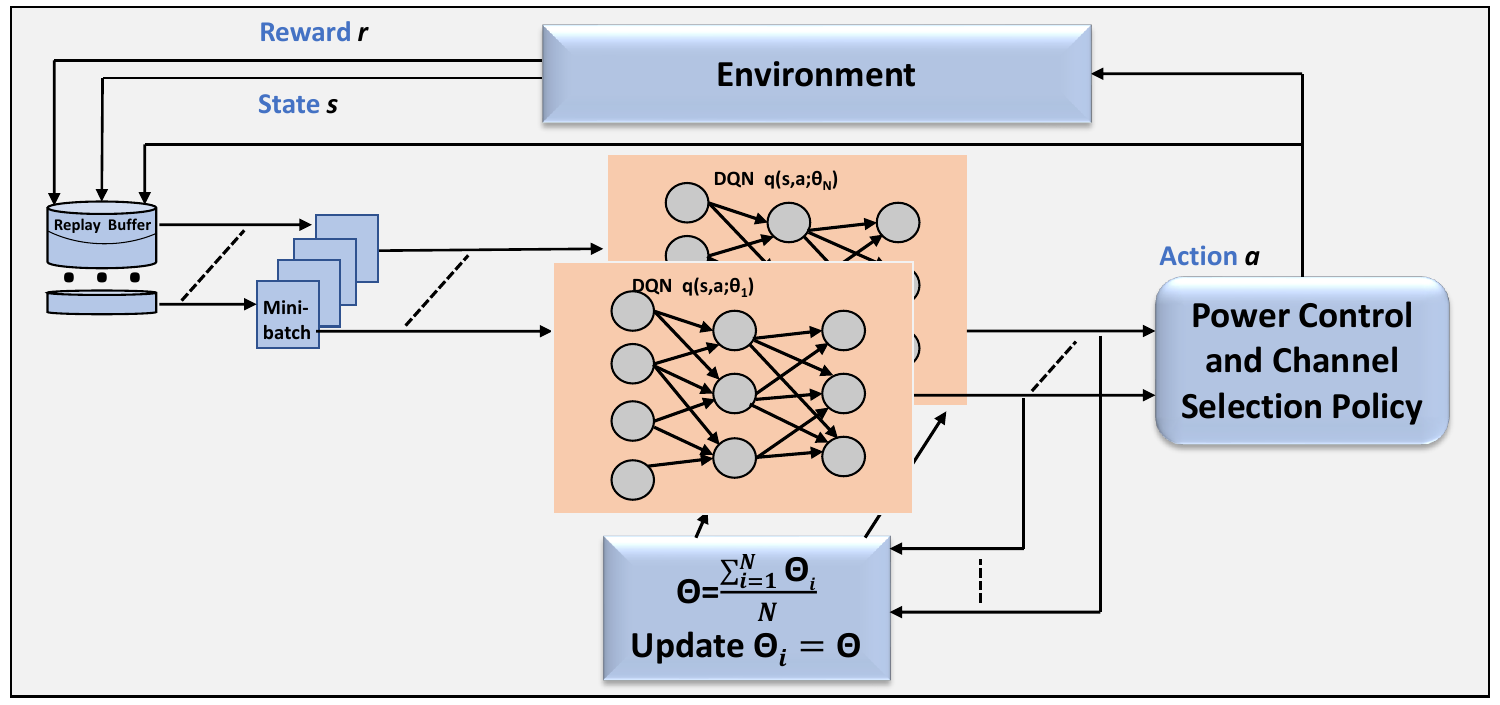} 
    	\caption{Multi-agent Deep Reinforcement Learning for the joint power control and channel selection } \label{fig:marlsetup} 
     \vspace{-0.2 in}
	\end{figure} 

%\subsection{Multi-Agent Power Control and Channel Selection Algorithm} 

We consider a problem of joint power control and channel selection in an active jamming environment through the lens of generalization. 
In this setup, each DRL agent optimizes its DQN parameters, $\boldsymbol\theta$, by the sampling data of minibatch size $B$ from the replay buffer $\mathcal{R}$ to optimize its policy using gradient descent algorithm. Further, agents in a network collaborate with other nodes to improve the generalization capacity, i.e., improve robustness in inference when deployed in different network operating scenarios.

\subsubsection{States}
Each Tx-Rx pair collects the local information and a few of the neighborhood information, which define the state of each node.  The state of node $i$ at the time slot $t$ can be expressed as $  \mathcal S_i^t = \big\{ \mathbb D_i^t, B_i^t, Ic_i^t, S_i^t, SINR^t \big\},
 $ where $\mathbb D_i^t = \{d^t_{ij}|j = 1,2,\cdots, K\}$ is the  set of distances to the neighboring receivers, $B_i^t$ is the number of packets in the buffer, $ Ic_i^t$ is the interference caused by transmission from node $i$ to the neighboring nodes, and  $S_i^t$ is the spectrum sensed by the transmitter node $i$. Note that the SINR is measured at the receiver and is communicated to the transmitter node $i$ through an ACK message. If the transmission from node $i$ is not received, it will not receive an ACK message, and the value for $S_i^t$ is set to be $-1$. Here, each agent maximizes its reward based on its state, which is a multi-agent reinforcement learning setup.    
  
  % Note that the dimension of the state of the node is a function of the number of nodes present on the network. One of the goals of the work is the deployment of the trained model in networks of different sizes. Hence, to make this a constant, we fix the value of $K$ and consider the distances of the closest receivers only if the number of nodes in a network is larger than $K$. Otherwise, we use the predefined largest value of distance if the number of nodes is less than $K$. 
   
   \subsubsection{Actions}
Let $p_1, \cdots, p_{n_p}$ be the power levels that agents can choose from and $f_1, \cdots, f_{n_f}$ be the number of frequencies that each agent can choose from. Then, the action of node $i$ at time slot $t$ is $
    \mathcal A_t^i = \big\{ (p^i,f^i) |  (p^i,f^i) \in \mathcal{A} \big \} $
Let $\mathcal P$ be the set of all available power levels, and $ \mathcal F$ be the set of all available frequencies, then the action space is  $    \mathcal A = \big\{ (p_i,f_i)| p_i \in \mathcal P \text{ and } f_i \in \mathcal F \big\}. $

\subsubsection{Rewards}
All the transmitters are required to transmit successfully while causing minimum interference to the neighboring node. Hence, it is desirable that each transmitter transmits successfully using the lowest power and choosing the available channels uniformly to minimize interference among nodes. 
To enforce all these constraints, we propose the following reward function for the given problem as 
\begin{align}
\begin{split}
    \mathcal R &= -C1 - C2,  \hspace{ 2mm}    \text{successful transmission,}
              \\ & = -C3,    \hspace{11mm}         \text{failure of the transmission, }
              \end{split}
\end{align}

{ \begin{algorithm}
\footnotesize
\caption{\small Generalized Training for DRL}
\label{alg1}
\begin{algorithmic}[1]
	\State{\textbf{Initialize:} DQN network parameters $\boldsymbol{\theta}_{i}^0 = \boldsymbol{\theta}^0, \forall i \in [{\color{blue}1:K}]$, learning rate $\alpha$}, Experience  replay buffer $\mathcal{R}$, and mini-batch ${B}$ 
	\For{Each Network configuration}
	\For{each episode $= 1,\cdots, N$}
    	\State{Observe an initial system state $s_t$}
    	\For{each time step $t=0, \cdots, T$}
    	\State{Select action $a_t$ at random with probability $\epsilon$ }
    	\State{Otherwise select $a_t$ as:$ a_t = \underset{a_t \in \mathcal{A}}{ \text{arg max}} ~~  q(s_t,a_t; \boldsymbol\theta_t) $ }
    	      \State {Execute action $a_t$, receive reward $r_t$ and state $s_{t+1}$}
    	      \State {Store the experience $e_t = (s_t, a_t,r_t, s_{t+1}$) }
    \EndFor
    \State{Update DQN parameter $\boldsymbol{\theta}$ using gradient-descent update}
    \State{Model Aggregation: $\bar{\boldsymbol{\theta}} = \frac{1}{K} \large\sum_{i=1}^ K \boldsymbol{\theta}_i$}, and update $ \boldsymbol{\theta}_i  = \bar{\boldsymbol{\theta}},  \forall i$
    % \State{Update DQN parameter  $ \boldsymbol{\theta}_i  = \bar{\boldsymbol{\theta}},  \forall i \in 1, \cdots, N $}
    \EndFor
    \State{Use Model Aggregation:  $\bar{\boldsymbol{\theta}} = \frac{1}{K} \large\sum_{i=1}^K \boldsymbol{\theta}_i$}
    , and update $ \boldsymbol{\theta}_i  = \bar{\boldsymbol{\theta}},  \forall i$.
    \EndFor
    \State{\textbf{Return}}
    
\end{algorithmic}
\end{algorithm}
}
\noindent where $C1$ refers to the cost of using power by an agent, $C2$ refers to the cost of switching frequencies, and $C3$ refers to the cost of failed transmission. 
% The choice of $C1$, $C2$, and $C3$ should be such that the successful transmission is power-efficient and the choice of frequencies is uniformly distributed.
The cost could be functions such as normalized SINR and normalized interference or constants. We choose $C1$ as $0.05$, $5$, and $10$ for using preset transmission power of $P_{low}$ dBm, $P_{med}$ dBm, and $P_{hig}$ dBm, respectively. In this work, we consider two frequencies available to each agent. 
To promote uniform distribution of choice of frequencies among agents, we create preference of channel uniformly among agents. 
We choose $C2$ as $0.05$ and $2$ for transmitting at the same channel and at a different channel. Note that the reward for transmitting with $P_{low}$ is higher than $P_{med}$, and the reward for transmitting with $P_{med}$ is greater than $P_{high}$ at any channel. Therefore, the choice of reward function makes agents prefer transmitting at a different channel to transmitting at higher power. Finally, we choose $C3$ as 20. These values are hyper-parameters and are chosen after numerous experiments. Let $r_t$ be the actual reward value obtained by an agent by selecting action $a_t \in \mathcal{A}$ when the agent is in the state $s_t \in \mathcal{S}$.

 \subsubsection{Training Strategies}
 The training should focus on the generalization of performance over wireless networks of different sizes and operational scenarios. Next, we enlist steps in the proposed training strategy,

 \begin{itemize}[leftmargin=0 pt]
    \item[]\emph{\textbf{States Uniformity:}} For this consideration, we have fixed the size of the state vector. The size of input for DQN for all the networks of different sizes will be the same and hence is the first step towards the generalization. In the state vector, the dimension of $\mathbb{D}$ depends on the size of the network. Instead, we fix the number $K$  and choose the $K$ most significant distances, i.e., the distances of the closest neighboring receivers where the interference caused by the agent is the most significant. 
    \item[]  \emph{\textbf{Node-level Aggregation:}}  The RF environment that each agent experience is limited. If the agents are trained individually, the policy learned by each agent is limited to the data seen by each agent. 
     When deployed, any of these agents will have worse performance as the training and testing scenario is going to be different. The problem is solved by node-level model aggregation. In the algorithm, we first train agents individually for a certain number of steps. We then aggregate the models and repeat the training. When the training converges, all the agents learn the policy corresponding to the RF scenario that all agents face in the network.
\begin{figure}
     \centering
     \subfloat[][{ Power Control}]{\includegraphics[width=.225\textwidth]{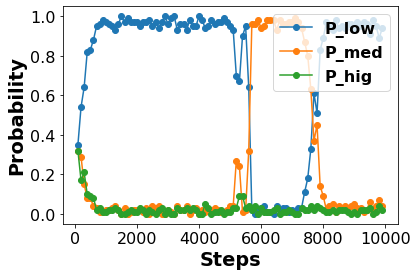}}
     \subfloat[][{ Channel Selection}]{\includegraphics[width=.225\textwidth]{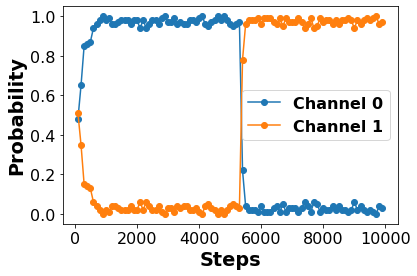}}
     \caption{Probabilities of Agent 1 as a function of steps (a) Power Control Probabilities (b) Channel Selection Probabilities.}\label{fig:agentActionSpace} 
     \vspace{-0.2in}
\end{figure}
     
%      \begin{figure}
%         \centering
%         % \includegraphics[width=\textwidth]{Full Length/Results/combined_PFprob.png}
%             \includegraphics[width=.5\textwidth]{Full Length/Results/n_b_Agent 1PowerFrequencyProbabilities.png}
%             % \caption{{\small Network 1}}    
%             \label{fig:actionAgent1}
%        % \begin{subfigure}[b]{1.1\textwidth}  
%         %     \centering 
%         %     \includegraphics[width=\textwidth]{Results/Agent 4Performance.png}
%         %     % \caption[]{{\small Network 2}}    
%         %     \label{fig:ActionAgent4}
%         % \end{subfigure} 
%         % \vspace{-.35 in}
%         \caption{ Power Control probabilities, and channel selection probabilities of Agent 1.} \label{fig:agentActionSpace}
%      %   \vspace{-.2 in}
% \end{figure}

\begin{figure}
     \centering
     \subfloat[][{ Training Performance}]{\includegraphics[width=.225\textwidth]{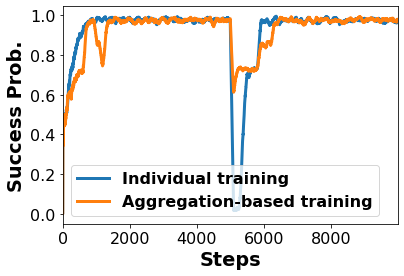}}
     \subfloat[][{ Testing Performance}]{\includegraphics[width=.225\textwidth]{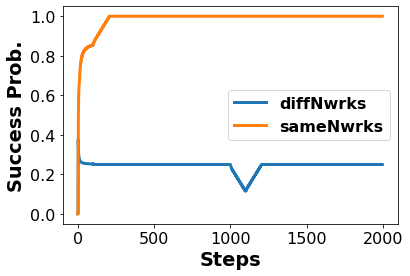}}
     \caption{Performance of Node-level Aggregation-based approach vs. individual training approach in (a) training and (b) testing (when deployed in a different network), respectively.}\label{fig:networktxprob} 
     \vspace{-.15 in}
\end{figure}

% \begin{figure}
%         \centering
%         % \begin{subfigure}[b]{0.235\textwidth}
%         %     \centering
%         %     \includegraphics[width=\textwidth]{Results/IndivAgg_Network_TX_success_prob.png}
%         %     \caption{{\small Training Comparison}}    
%         %     \label{fig:mean and std of net14}
%         % \end{subfigure}
%        \centering 
%         \includegraphics[width=.48\textwidth]{Full Length/Results/n_b_IndpTraDiffTest.png}
%          \caption{Performance comparison of Node-level Aggregation-based approach (with individual training approach) in training and testing (when deployed in a different network), respectively.}\label{fig:networktxprob} %\vspace{-.3 in }
% \end{figure} 

% \begin{figure}
%      \centering
%      \begin{subfigure}[b]{0.24\textwidth}
%          \centering
%          \includegraphics[width=\textwidth]{Full Length/Results/n_b_IndpTraDiffTest.png}
%          \caption{$y=x$}
%          \label{fig:y equals x}
%      \end{subfigure}
%      \hfill
%      \begin{subfigure}[b]{0.24\textwidth}
%          \centering
%          \includegraphics[width=\textwidth]{Full Length/Results/n_b_IndpTraDiffTest.png}
%          \caption{$y=3\sin x$}
%          \label{fig:three sin x}
%      \end{subfigure}
%      % \hfill
%      % \begin{subfigure}[b]{0.3\textwidth}
%      %     \centering
%      %     \includegraphics[width=\textwidth]{graph3}
%      %     \caption{$y=5/x$}
%      %     \label{fig:five over x}
%      % \end{subfigure}
%      %    \caption{Three simple graphs}
%      %    \label{fig:three graphs}
% \end{figure}

     \item[]  \emph{\textbf{Network-level Aggregation:}} Further, we should emphasize that creating all possible RF scenarios within a network is difficult. First, the network grows very large, and second, creating individual RF scenarios across each node in a bigger network is difficult compared to a small network. We further extend the ideas of model aggregation across nodes to aggregation over networks. Here, we design several smaller networks, where each network has specific network scenarios. Next, we train the agents of each network until convergence individually and save the DQN parameters. We aggregate all these DQN parameters from each network scenario and use the average DQN as the starting network in each node. Starting with the averaged model, we iterate over each network with model aggregation in between until convergence in all network scenarios. This approach ensures each agent learns a policy that can accommodate all wireless environments seen by agents in all networks. 
     % In addition, the training results of all previous experiments conducted are reusable. We can save the trained model and use it again if necessary. At the same time, we can easily add or remove scenarios by adding or removing models during aggregation. 
     
     % We do not have to go over the entire training process again. We need to aggregate the model and train them together.
     \item[] \emph{\textbf{Introducing Randomness:}} 
     % Even though several networks are considered for training purposes, the agents only learn as many scenarios as these agents face, which is again, a fixed number of scenarios.
     It is a common approach in DRL to train and test the learned agent in the same environment. However,  wireless networks tend to be different from the training network during deployment. While network-level aggregation helps to address some of the variations in the wireless environments, introducing stochasticity in the environment itself, such as by using a random walk-based mobility model, improves the generalization capacity of the policy of DRL agents to small variations in wireless networks. 
     
     The generalized training algorithm is summarized in \ref{alg1}.
      \end{itemize} 

 % With all these measures, we train DQN agents for generalized training. 
 % Th/e following section discusses the training and inference results we get from the proposed approach.
%  \begin{figure}
%         \centering
%         \begin{subfigure}[b]{0.240\textwidth}  
%             \centering 
%             \includegraphics[width=\textwidth]{Results/SamevsDiffNtwk.png}
%             \label{fig:DiffNtwk}
%         \end{subfigure}
%         \caption{\small Transmission success probability on an unseen network } \label{fig:txprobTest}
% \end{figure}  
 
%
 \section{Experiment Results}
For the experiments, we use MR-iNet Gym \cite{FARQUHAR2022} that uses ns3-gym \cite{gawlowicz2019ns}. This includes our custom DS-CDMA module for ns-3 to simulate a distributed LPI/D wireless network controlled by DRL running in an OpenAI-Gym. All training networks were developed using this framework. For each network trained and tested, we had jammers disrupting the transmission. The jammer was switched on at the same time when the Tx-Rx nodes were active. Further, for every training experiment, the jammer  changed jammed frequency during the course of the training. Each agent uses the same fully connected network as the Q-network, which has  hidden layers with  128 and 256 nodes, respectively. We use ReLU as the non-linear function in the hidden layers nodes and linear activation in the output layer nodes. Adam optimizer is used for the evaluation of stochastic gradient descent with a learning rate of 0.0001.
%  \vspace{-.2 in}

%  \begin{figure*}[t]
% \centering
%         \begin{subfigure}[b]{0.26\textwidth}
%             \centering
%             \includegraphics[width=\textwidth]{Full Length/Results/bold_Power6.png}
%             \end{subfigure}
%         \hfill
%         \begin{subfigure}[b]{0.26\textwidth}  
%             \centering 
%             \includegraphics[width=\textwidth]{Full Length/Results/bold_Frequency6.png}
%            \end{subfigure}
%         \hfill
%         \begin{subfigure}[b]{0.26\textwidth}
%             \centering
%             \includegraphics[width=\textwidth]{Results/bold_Network_TX_success_prob.png}
%            \end{subfigure}
%         \hfill
%         \vspace{-.05 in}
%         \caption{\small Power Control, Channel Selection, and Transmission Success Probability with aggregation among networks, respectively.} 
%         \label{fig:txsuccesstraintest}
% 		\label{fig:multiNet} 
%   \vspace{-.2 in}
% \end{figure*}
\begin{figure*}
     \centering
     \subfloat[][{ Power Control}]{\includegraphics[width=.3\textwidth]{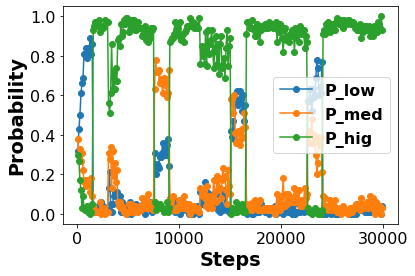}}
     \subfloat[][{ Channel Selection}]{\includegraphics[width=.3\textwidth]{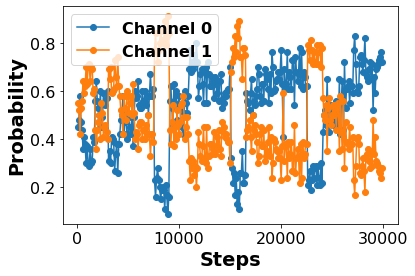}}
     \subfloat[][{ Tx. Success Probability}]{\includegraphics[width=.3\textwidth]{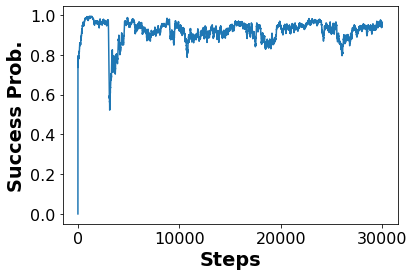}}
     \caption{Training performance of the proposed generalized approach as a function of steps: (a) Power Control Probabilities, (b) Channel Selection Probabilities, and (c) Transmission Success Probability with aggregation among networks, respectively.}\label{fig:trainingPlots}
     \vspace{-0.1 in}
\end{figure*}

%  \begin{figure*} 
%   \centering
%     	\includegraphics[width= 15.5cm, height= 3.15 cm]{Full Length/Results/n_b_trainingResult_2.png} 
%     	\caption{\vspace{-0.05 in} Power Control, Channel Selection, and Transmission Success Probability with aggregation among networks, respectively. } \label{fig:trainingPlots} %\vspace{-0.2 in}
% \end{figure*} 

\begin{figure}
     \centering
     \vspace{-0.27 in}
     \subfloat[][{ Cumulative Reward}]{\includegraphics[width=.25\textwidth]{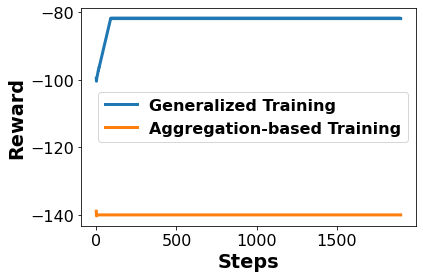}}
     \subfloat[][{ Tx. Success Probability}]{\includegraphics[width=.25\textwidth]{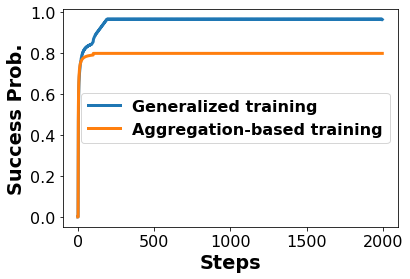}}
    \caption{ Performance comparison of agents when tested on an unseen network of size 10 using aggregation-based training and with proposed training approach: (a) Cumulative Reward; and (b) Network Transmission Success Probability.} 
        \label{fig:txsuccessdiffArch}
\end{figure}

 % \begin{figure}
 %        % \centering
        
 %           \includegraphics[width=0.5\textwidth]{Full Length/Results/b_n_test10nodes.png}
           
 %        \caption{ Cumulative Reward and Network Transmission Success Probability of agents when tested on an unseen network of size 10 using aggregation-based training and with proposed strategy, respectively.} 
 %        \label{fig:txsuccessdiffArch}
 %        %\vspace{-.15 in}
 %    \end{figure}

  \subsection{Simulation-based Performance}

\begin{figure}
         \centering
        
           \includegraphics[width=0.275\textwidth]{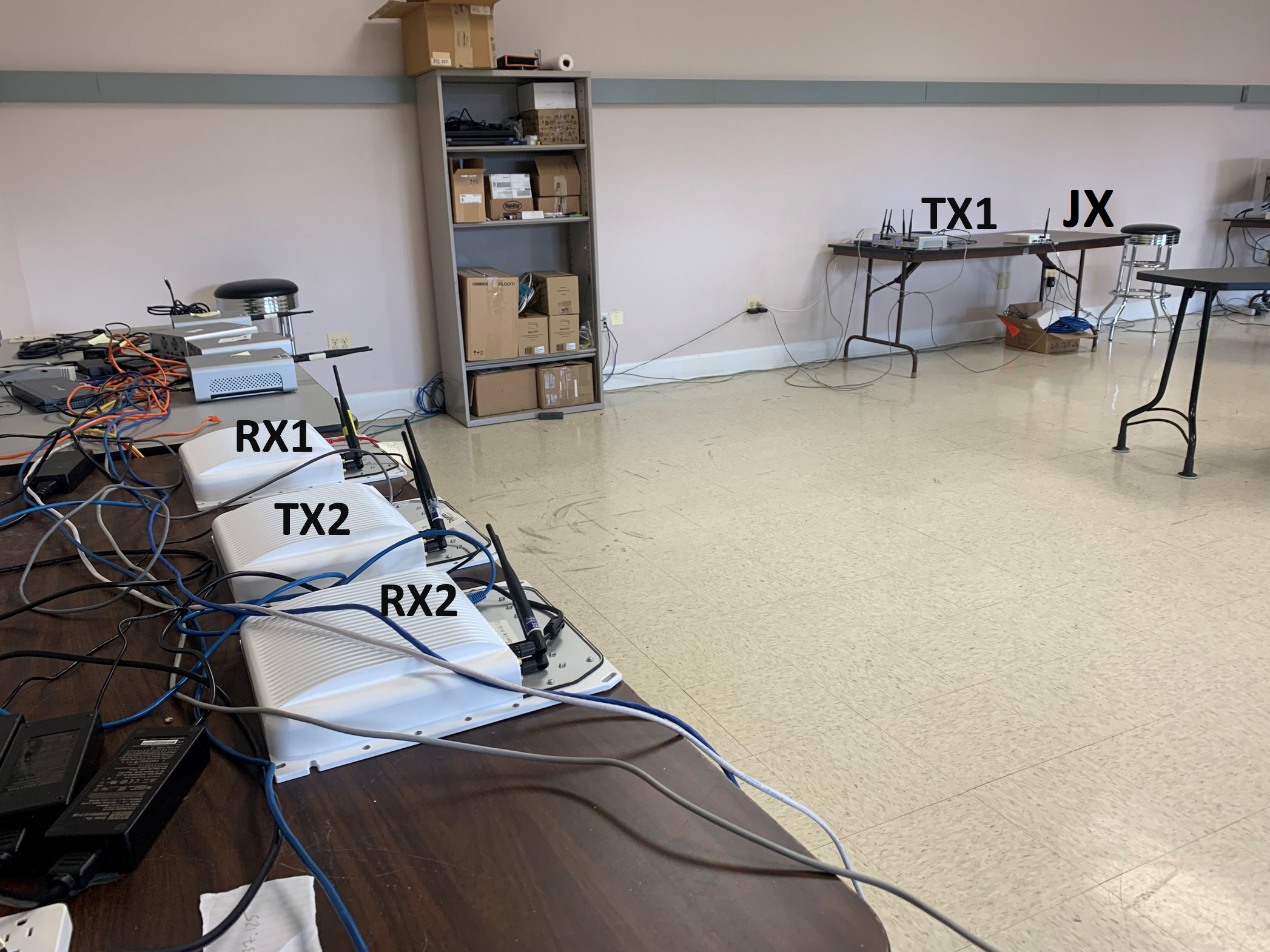}
  \caption{In-lab Testbed }
  \label{fig:testbed}
  \vspace{-.2 in}
  \end{figure}

% \begin{figure}[t!]
% \centering
% \begin{minipage}{.24\textwidth}
%   \centering
%   \includegraphics[width=4.20 cm]{Full Length/Testbed.png}
%   \captionof{figure}{In-lab Testbed }
%   \label{fig:testbed}
% \end{minipage}\hspace{2 pt}%
% \begin{minipage}{.24\textwidth}
%   \centering
%   \includegraphics[width=4.20 cm, height = 2.9 cm]{Full Length/Scenario3.png}
%   \captionof{figure}{Outdoor Test Scenario: Three pairs with two jammers}
%   \label{fig:outdoorTest}
% \end{minipage}
% \vspace{-0.2 in}
% \end{figure}
  
  In the first experiment, we compare the training performance of the node-level aggregation-based training approach with individual training. Each Tx-Rx pair can communicate using three power levels in two different channels. Jammers disrupt communication between radios by changing their jamming frequency in the middle of the training. In Fig. \ref{fig:agentActionSpace}, we plot Agent-1's power and channel selection probabilities, where we can see agents deciding to choose minimum power and avoid jamming frequency to get successful transmission. In Fig. \ref{fig:networktxprob}, we can see that while the node-level aggregation-based approach learns almost as good as the individual training approach, the node-level aggregation-based approach has better performance when there is a change in network dynamics, i.e., jamming frequency. However, from Fig. \ref{fig:networktxprob}, we can see that the test performance of the node-level aggregation-based approach in an unseen network deteriorates, which motivates the need for the next part of our approach.

  In the second experiment, we consider the proposed training approach, which also included network-level aggregation. In this setup, we consider five networks so that each agent learns a policy for different wireless scenarios. 
   Agents are trained separately for each network, and the DQN model is saved. We then aggregate all these saved models and retrain agents with node-level aggregation for each network sequentially until we exhaust all networks. We repeat this training until convergence. Next, we plot the training results of this experiment in Fig. \ref{fig:trainingPlots}. We can see that the agent is changing power and frequencies with time depending on which network is being switched on. We should note that the network transmission success probability is approaching unity with increasing training steps. This suggests that each node is converging to a policy that can address all possible scenarios encountered by all agents in different networks.
  Next, we used this trained agent in an \textbf{\textit{unseen network}} of size 10. The results of these tests are shown in Fig. \ref{fig:txsuccessdiffArch}.
  % We plot the results of the model-aggregation-based approach and the proposed methods in Figure \ref{fig:txsuccsstest_ind10}. 
  From the results, we can see from the inference that the proposed method outperforms the node-level aggregation-based approach in both cumulative rewards and network transmission success probability, demonstrating the proposed approach's \textbf{\textit{effectiveness in an unseen network.}}

\begin{table}
\centering
\small
\begin{tabular}{ |p{1cm}||p{1.5cm}|p{1cm}|p{1cm}| p{1cm}| }
\hline
 Nwk & Parameters &PA & Rd & Gd\\
 \hline
    & PDR &89.88    & 45.33 &   75.71\\
 Nwk 1  &  SE & 3.17  & 0.48   & 3.01\\
 \hline
  & PDR & 91.23 & 48.88 &  72.45\\
 Nwk 2  &  SE    &3.15 & 0.51&  3.02\\
 \hline
    & PDR &   90.63 & 47.69 & 76.13\\
Nwk 3&  SE & 3.15  & 0.48   &3.01\\
\hline
\end{tabular}
\caption{PDR and SE of the proposed algorithms (PA), Gd, Rd in three different network scenarios.}\label{table:OTA}
\vspace {-.25 in}
\end{table}
\vspace{-.1 in}
\subsection{Over-the-air Hardware Experiments and Demonstration}
 %\vspace{-.1 in}
In the final experiment, we conduct an OTA experiment where we deploy the inference engine obtained through the generalized training approach in the MR-iNet Gym environment. The DRL inference engine is now compressed using TensorRT, which not only reduces the model size of the inference engine but also reduces its processing time. In this experiment, we take two pairs of SDRs acting as two Tx-Rx pairs in an active jamming environment. The jammer changes the transmission frequency at random every 10 seconds to disrupt the communication between the Tx-Rx pairs. 

In this setup, the Tx-Rx pairs are set to transmit within ISM band (S-Band). The transmitters are set to transmit at three different power levels. 
We transmit video packets over the SDRs, measure packet delivery rate (PDR) and spectral efficiency (SE), and compare the performance of the proposed algorithm to two different algorithms. The first algorithm selects each of the available frequencies and power options available to the transmitter at random. We refer to this approach as \texttt{Random} (Rd), which is similar to the baseline used in \cite{nasir2021deep}. This serves as the lower baseline for the proposed algorithm. The second algorithm scans the available channel and transmits it on the channel with the least interference with the maximum power, which is referred to as \texttt{Greedy} (Gd). A single topology of the in-lab hardware testbed is shown in Fig. \ref{fig:testbed}. The results in Table \ref{table:OTA} shows that the PDR and SE of the proposed algorithm are consistently better than the other approaches in all networks. 
The proposed algorithm makes better power control and channel selection decisions than other algorithms such that most packets are transmitted successfully in all the radios. 
In addition, all the networks considered for OTA evaluation \textbf{\textit{were different from the networks used for training}}, which shows that the inference engines obtained through the proposed training approach in the ns3-gym environment generalize over the unseen wireless networks.

\begin{table}
\centering
\small
\begin{tabular}{ |p{1.0cm}||p{1.5cm}|p{1cm}|p{1cm}| p{1cm}| }
\hline
 Jammer & Parameters &PA & Rd & Gd\\
 \hline
    & PDR &97.81    & 54.71 &   88.22\\
 CJ  &  SE & 3.47  & 0.85   & 3.15\\
 \hline
  & PDR & 97.79 & 48.88 &  72.45\\
 RHJ  &  SE    &3.46 & 0.84&  3.14\\
 \hline
    & PDR &   97.89 & 47.69 & 76.13\\
ChJ &  SE & 3.47 & 0.84   &3.14\\
\hline
\end{tabular}
\caption{ PDR and SE of PA, Gd, and Rd in three different jamming scenarios:a) constant jammer (CJ), b) Random Hopping Jammer (RHJ), and c) Chirp Jammer (ChJ)}\label{table:Jammer}
\vspace {-.2 in}
\end{table}

Next, we compare the performance of the proposed algorithm in three different jamming environments; (i) Constant Jammer (CJ) is a Band-limited Gaussian jammer, which jams one of the channels that radios use for communication, (ii) Random Hopping Gaussian Jammer (RHJ), which randomly hops between the available channels. (iii) Chirp Jammer (ChJ), which sweeps across all frequencies in the signal bandwidth. We use a chirp period of $\frac{1}{4}$ and a chirp rate of 4.8 MHz/sec. The Chirp jammer also hops between the available channels. The results of this comparison are listed in Table \ref{table:Jammer}. From the results, we can see that the DRL agent avoids the jammer in all three scenarios and hence has a very high PDR in all three jamming scenarios. Further, SE is also comparable in all three scenarios.  

\begin{figure}
        \centering
           \includegraphics[width=0.3\textwidth]{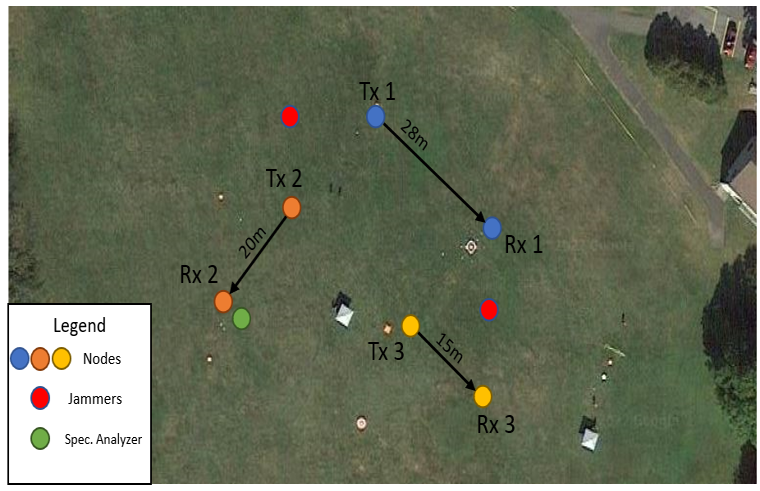}
  \caption{Outdoor Test Scenario: Three pairs with two jammers }
 \label{fig:outdoorTest}
 \vspace {-.25 in}
  \end{figure}

Finally, we deployed the DRL engine on three pairs of radios with two jammers in an outdoor real-world environment. The deployment setup is shown in Fig. \ref{fig:outdoorTest} with the demonstration video of the real-time operation of the DRL inference engine available in \cite{video2} with a voice-over description.
%\vspace{-.1 in}
 \section{Conclusion}
% We presented our generalized algorithmic approach to training agents for DRL problems in wireless networks. We demonstrated the effectiveness of our approach using the problem of distributed joint control and channel selection in a hostile jamming environment. Unlike prior works that focused on specific network architecture, we generalize the training performance that shows good training and testing performance across seen and unseen networks. The proposed algorithm demonstrates superior power control and channel selection action with improved network metrics, such as transmission success probability in both simulated environments and actual wireless networks created using SDRs. Finally, we demonstrated how the solution was implemented on the radios real-time operation in an outdoor real-world scenario. 
We investigated the problem of joint power control and frequency selection among multiple Tx-Rx pairs in a hostile jamming environment with the goal of hardware deployment. Striving for the real-world operating scenarios of the trained agent in unseen networks, we presented our generalized algorithmic approach to training agents for DRL problems in wireless networks. Unlike prior works that presented results on a specific network architecture, the proposed approach showed improved training and testing performance across seen and unseen networks for the problem of joint power control and frequency selection in both simulated and actual wireless networks. Finally, we demonstrated the video streaming operation of the agent in an outdoor hostile jamming environment to prove the impact of the proposed solutions. We hope the proposed approach can accelerate the deployment of DRLs for real-world wireless networks.

\bibliography{wireless_RL}

\bibliographystyle{IEEEtran}
%\let\clearpage\relax

%\par\leavevmode

\end{document}